# Study of Total, Absorption, and $^3$He and $^3$H Production Cross Sections in $^4$He-proton Collisions

Congchong Yan[1], Premkumar B. Saganti[2]

Francis A. Cucinotta[3,*]

[1]School of Radiation Medicine and Protection, Institute of Space Life Sciences, Medical College of Soochow University, Suzhou, China

[2]Physics Department, Prairie View A&M University, Prairie View TX, United States of America

[3]Department of Health Physics and Diagnostic Sciences, University of Nevada, Las Vegas, NV, United States of America

*Correspondence author
E-mail: francis.cucinotta@unlv.edu

Keywords:   cosmic rays, $^4$He absorption cross sections, $^4$He fragmentation, Glauber model

**Abstract**

Light ion breakup cross sections are important for studies of cosmic ray interactions in the interstellar medium or radiation protection considerations of energy deposition in shielding and tissues. Abrasion cross sections for heavy ion reactions have been modeled using the Glauber model in the large mass limit or Eikonal form of the optical potential model. Here we formulate an abrasion model for $^4$He fragmentation on protons using the Glauber model avoiding the large mass limit and include a model for final state interactions. Calculations of energy dependent total, absorption, elastic and breakup cross sections for $^4$He into $^3$He or $^3$H with protons targets are shown to be in good agreement with experiments for energies from 100 to 100,000 MeV/u. The Glauber model for light nuclei with and without a large mass limit approximation are shown to be in fair agreement above 300 MeV/u, however important differences occur at lower energies.

## 1. Introduction

The $^4$He nucleus is the second most abundant cosmic ray nuclei after protons. Studies of interaction cross sections of $^4$He with hydrogen or heavier atoms are important for understanding inter-stellar cosmic rays transport, including understanding source particle spectra [1-3], and for radiation protection in spaceflight [4-6]. Previously we considered a theoretical model based on a cluster expansion of the internal $^4$He structure in interactions with target nuclei and final state interactions [7-9]. In this model $^4$He internal cluster interactions with target nuclei were modeled using the Glauber model for elastic and inelastic scattering in nucleon-nucleus [10] or composite cluster-nucleus scattering using realistic internal wave functions to represent vertex functions for nuclear dissociation [7-9]. On the other hand, the abrasion-ablation model of heavy ion fragmentation has been shown to be fairly accurate with the abrasion stage modeled using the Glauber [11] or optical model in the Eikonal approximation [12-14], and ablation described by the statistical decay model. Excitation energies are evaluated using several approaches including center-of-mass distortions and final state interactions of abraded nucleons with a pre-fragment. However this model has not been applied to 4He fragmentation in the past.

Because of the success of the abrasion model for fragmentation of $^{12}$C or heavier projectiles we decided to test this model for describing interaction cross sections for the important light nucleus, $^4$He. In this paper we consider the abrasion model applied to $^4$He interactions with protons (hydrogen targets). The focus of our previous model [7-9] was double differential cross sections in angle and momentum, which are not considered in the present work, and instead we consider energy dependent integrated cross sections for beam energies from 100 to 100,000 MeV/u. Using the independent particle model (IPM) applied to the Glauber model we compare calculations with and without the large mass limit approximation. Final state interactions of knocked out protons or neutrons are modeled using a classical approach [15]. We include comparisons to more recent experimental data since our earlier work. Predictions of energy dependent cross sections are shown to be in fairly good agreement with experiments found.

## 2. Application of Glauber Model for Light-Ion Abrasion

### 2.1 Glauber Model Formalism

The Glauber scattering operator is defined in terms of the nuclear profile function as [9]:

$$\hat{f}(\vec{q}) = \frac{ik}{2\pi} \int d\vec{b} \exp(i\vec{q}\cdot\vec{b})\hat{\Gamma}(\vec{b}) \qquad (1)$$

where $q$ is the momentum transfer, $b$ is the impact parameter, and $k$ the relative projectile-target momentum in the center of mass frame. The nuclear profile function is a found as a product of the nucleon-nucleon profile functions:

$$\hat{\Gamma}(\vec{b}) = 1 - \prod_{j}^{A}[1 - \Gamma_j(\vec{b} - \vec{s}_j)] \qquad (2)$$

We consider a light projectile nucleus interaction with protons (hydrogen targets) with $A$ the projectile mass number, with $\vec{s}_j$ the transverse component and $z_j$, the longitudinal components of the internal coordinates, $\vec{r}_j = (\vec{s}_j, z_j)$. Using the IPM for the projectile wave function the elastic amplitude is found as [9]:

$$f_{elas}(q) = \frac{ik}{2\pi} \int d\vec{b}\, e^{i\vec{q}\cdot\vec{b}} \left\{ 1 - \prod_{j=1}^{A}[1 - \int d\vec{q}'\, e^{-i\vec{q}'\cdot\vec{b}} f_{NN}(\vec{q}')F(\vec{q}')] \right\} \quad (4)$$

where $f_{NN}(q)$ is the nucleon-nucleon scattering amplitude and $F(q)$ the projectile one-body ground state form factor. The total scattering cross section is evaluated using the optical theorem and equation (4), which leads to:

$$\sigma_{Tot} = 4\pi \int b\, db\, [1 - \mathrm{Im} \prod_{j=1}^{A}(1 - M_j(\vec{b}))] \quad (5)$$

with $M_j(b)$ defined below. Using equation (4) the total elastic cross section is found after integrating over the solid angle using $d\Omega \sim d^2q/k^2$ as:

$$\sigma_{Elas} = 2\pi \int b\, db \left| 1 - \prod_{j=1}^{A}(1 - M_j(\vec{b})) \right|^2 \quad (6)$$

The absorption (total inelastic) cross section can then be evaluated using $\sigma_{Tot} - \sigma_{Elas}$ from equations (5) and (6), however it is useful to calculate this cross-section more directly. An alternative approach to evaluate the absorption cross section is to use closure over the projectile final states in the IPM [9], which leads to

$$\frac{d\sigma_{Abs}}{d\Omega}(\vec{q}) = \left(\frac{k}{2\pi}\right)^2 \int d\vec{b}\, d\vec{b}'\, e^{i\vec{q}\cdot(\vec{b}-\vec{b}')} \left\{ \begin{array}{l} \prod_{j=1}^{A}[1 - M_j(\vec{b}) - M_j^\dagger(\vec{b}') + \Omega_j(\vec{b},\vec{b}')] - \\ \prod_{j=1}^{A}[1 - M_j(\vec{b})][1 - M_j^\dagger(\vec{b}')] \end{array} \right\} \quad (7)$$

The functions $M_j(\vec{b})$ and $\Omega_j(\vec{b},\vec{b}')$ are defined using single particle densities, $\rho(r)$, or form-factors, $F(q)$, by,

$$M_j(\vec{b}) = \int d\vec{r}_j\, \Gamma_{NN}(\vec{b} - \vec{s}_j) \rho_j(\vec{r}_j) \quad (8)$$

or equivalently,

$$M_j(\vec{b}) = \frac{1}{2\pi i k_{NN}} \int d\vec{q}\, e^{-i\vec{q}\cdot\vec{b}} f_{NN}(\vec{q}) F_j(\vec{q})] \quad (8')$$

and

$$\Omega_j(\vec{b},\vec{b}') = \int d\vec{r}_j \Gamma_{NN}(\vec{b}-\vec{s}_j)\Gamma^\dagger_{NN}(\vec{b}'-\vec{s}_j)\rho_j(\vec{r}_j) \qquad (9)$$

or equivalently,

$$\Omega_j(\vec{b},\vec{b}') = \frac{1}{(2\pi k_{NN})^2}\int d\vec{q}d\vec{q}'\, e^{i\vec{q}\cdot\vec{b}}\, e^{i\vec{q}'\cdot\vec{b}'} f_{NN}(\vec{q})f^\dagger_{NN}(\vec{q}')F(\vec{q}+\vec{q}') \qquad (9')$$

Elastic scattering dominates at small momentum transfer, $q$. Away from the forward scattering region equation (7) can be approximated by the following factorized form [9]:

$$\frac{d\sigma_{Abs}}{d\Omega}(\vec{q}) = (\frac{k}{2\pi})^2 \int d\vec{b}d\vec{b}'\, e^{i\vec{q}\cdot(\vec{b}-\vec{b}')} \prod_{j=1}^{A}[1-M_j(\vec{b})][1-M^\dagger_j(\vec{b}')]\prod_{j=1}^{A}[1+\Omega(\vec{b},\vec{b}')-1] \qquad (10)$$

After integration over all momentum transfers, the total inelastic (absorption) cross section is found as:

$$\sigma_{Abs} = \int d\vec{b} \prod_{j=1}^{A}\left|1-M_j(\vec{b})\right|^2 \left\{\prod_{j=1}^{A}[1+\Omega_j(\vec{b},\vec{b})]-1\right\} \qquad (11)$$

The fulfillment of the unitarity condition in the Glauber model [10, 16-18] has been discussed over the years and leads to the following condition on the NN profile function:

$$\Gamma_{NN}(\vec{b})\Gamma^\dagger_{NN}(\vec{b}) = \Gamma_{NN}(\vec{b}) + \Gamma^\dagger_{NN}(\vec{b}) \qquad (12)$$

We found that it is more accurate to use equation (12) in the evaluation of equation (9) when $\vec{b}=\vec{b}'$. Here we note that the $\Omega$ function is defined in terms of a two-particle operator that has a much larger number of correction terms to the IPM for two-body correlations [19] compared to the $M(b)$ function which involves a one-body operator. Use of equation (12) allows us to recast equation (11) as:

$$\sigma_{Abs} = \int d\vec{b} \prod_{j=1}^{A}\left|1-M_j(\vec{b})\right|^2 \left\{\prod_{j=1}^{A}[1+M_j(\vec{b})+M^\dagger_j(\vec{b})]-1\right\} \qquad (13)$$

Equation (13) is found to be more advantageous and accurate in numerical calculations than equation (11).

Following arguments made for heavy ion abrasion [12-14] we write the abrasion cross section for removal of $n$ nucleons from the projectile nucleus:

$$\sigma_{abr}(n) = \binom{A}{n}\int d\vec{b}\prod_{j=1}^{A-n}\left|1-M_j(\vec{b})\right|^2 \left\{\prod_{j=1}^{n}[1+M_j(\vec{b})+M^\dagger_j(\vec{b})]-1\right\} \qquad (14)$$

In equation (14) the first product represents the spectator term of nucleon interactions with the target, while the second-factor represents the participant nucleons that undergo particle knockout or so-called abrasion. The large A or optical model limit is then found from equation (14) as A>>1 to be identical to the optical model of abrasion.

## 2.2. Final State Interactions

We use the model of Abul-Magd and Hufner [15] to describe the final state interaction of knocked-out neutron or proton with a $^3$He or $^3$H nucleus, respectively. In this model the knocked-out nucleon deposits energy in the fragment before it leaves the interaction region. The nucleon moves in the impact parameter plane. Its trajectory for a given impact parameter, $b$, is then described by its azimuthal scattering angle, $\varphi$, for a fragment with radii, $R$,

$$L(\varphi) = \sqrt{R - b\sin^2(\varphi)} + R\cos(\varphi) \qquad (15)$$

The energy loss is described by

$$\frac{d\varepsilon}{dL} = -\frac{\alpha_L}{\lambda}\varepsilon \qquad (16)$$

where $\lambda$ is the mean free path and $\alpha_L$ is the energy loss per collision. The knocked-out nucleon energy is not described but assumed to have a relative motion with the fragment with kinetic energy of a few 10's of MeV, where isotropic scattering approximately holds such that $\alpha_L \sim 0.25$.

The distribution of excitation energies is then described by the following [15],

$$S(E) = \int_0^\infty d\varepsilon W(\varepsilon) \int_0^{2\pi} \frac{d\varphi}{2\pi} \delta[E - \varepsilon(1 - \exp(-\varepsilon_L L(\varphi)/\lambda)] \qquad (17)$$

In equation (17), $W(\varepsilon)$ is the distribution of initial energies of the knocked-out nucleon in the fragment rest frame, which is assumed to be exponential with a mean of about 80 MeV [15].

For the light fragments $^3$He and $^3$H, energy deposition above ~8 MeV will lead to removal of an additional nucleon leading to $^2$H and nucleons. We therefore can reduce the fragment cross sections by the fraction of energy deposited above 8 MeV, S(E>8 MeV)/S(E>0 MeV) to obtain the final $^3$He and $^3$H fragment cross sections. This ignores Coulomb differences and the small difference in nucleon separation energies between $^3$He and $^3$H.

## 2.3. Physical Inputs

For physical inputs we use the diffractive form of the NN scattering amplitude:

$$f_{NN}(q) = \frac{(\alpha + i)k_{NN}\sigma_{NN}}{4\pi} \exp(-\frac{1}{2}B_{NN}q^2) \qquad (11)$$

where $\sigma_{NN}$ is the NN total cross section, $\alpha_{NN}$ the ratio of the real to imaginary part of the forward amplitude and $B_{NN}$ the slope parameter. The total neutron-proton cross section is parameterized to experimental data [21] with kinetic energy $T$ in GeV (see also **Figure 1**) as:

$$\sigma_{np} = \exp\left(\frac{p_1 T^4 + p_2 T^3 + p_3 T^2 + p_4 T + p_5}{T^4 + q_1 T^3 + q_2 T^2 + q_3 T + q_4}\right)$$

$$T = \frac{T_{Lab} - 1.922}{2.839}$$

$$p1 = 4.025, p2 = 22.43, p3 = 48.04, p4 = 44.26, p5 = 15.19$$

$$q1 = 5.967, q2 = 13.18, q3 = 12.17, q4 = 4.132$$

The proton-proton cross section is parameterized as (**Figure 1**):

For $T_{Lab} < 0.3$ GeV

$$\sigma_{pp} = \frac{p_{L1}T^2 + p_{L2}T + p_{L3}}{T^2 + q_{L1}T + q_{L2}}$$

$$p_{L1} = 38.8, p_{L2} = -1.678, p_{L3} = 0.7487,$$
$$q_{L1} = 0.2412, q_{L2} = -2.706 \times 10^{-4}$$

For $T_{Lab} \geq 0.3$ GeV

$$\sigma_{pp} = \exp\left(\frac{p_{H1}T^4 + p_{H2}T^3 + p_{H3}T^2 + p_{H4}T + p_{H5}}{T^4 + q_{H1}T^3 + q_{H2}T^2 + q_{H3}T + q_{H4}}\right)$$

$$T = \frac{T_{Lab} - 1.922}{2.839}$$

$$p_{H1} = 4.025, p_{H2} = 22.43, p_{H3} = 48.04, p_{H4} = 44.26, p_{H5} = 15.19$$

$$q_{H1} = 5.967, q_{H2} = 13.18, q_{H3} = 12.17, q_{H4} = 4.132$$

For the ratio of the real to imaginary part of the forward NN amplitude we fit the following functions to experimental data [21] or the estimates by Grein [22] (also see **Figure 1**):

For $T_{Lab} < 1$ GeV

$$\alpha_{np} = \frac{p_1 T^2 + p_2 T + p_3}{T^3 + q_1 T^2 + q_2 T + q_3}$$

$$T = \frac{T_{Lab} - 0.1857}{0.2556}$$

$$p1 = -4.473, p2 = -0.5982, p3 = 1.649, q1 = 4.531, q2 = 5.058, q3 = 2.495$$

For $T_{Lab} \geq 1$ GeV

$$\alpha_{np} = 0.0865 \ln(T_{Lab}) - 0.4712$$

$$\alpha_{pp} = \begin{cases} -3.2263 \exp(-31.56 T_{Lab}) + 3.72875 \exp(-4.683 T_{Lab}) - 0.5 & T_{Lab} < 2\ GeV \\ 0.0963 \ln(T_{Lab}) - 0.5265 & T_{Lab} \geq 2 GeV \end{cases}$$

For the energy dependent slope parameters with B in units of fm$^{-1}$, we use the fits from [20]

$$B_{np} = 0.0788[3.5 + 30 \exp(-T_{lab}/200)]$$

$$B_{pp} = 0.0788[3. + 14 \exp(-T_{lab}/200)]$$

The energy dependent parameters in equation (14) consider proton and neutron values corresponding to removal of neutrons or protons from the projectile in interactions with the proton target.

The Glauber model is usually found to be accurate above a few hundred MeV/u. To increase the accuracy at lower energies (here considering as low as 100 MeV/u), a medium modified NN cross section can be used. The models in the literature show some variation in estimating medium effects [22,23], however converging to the free NN cross section at about ~300 MeV. The medium modification in the 100 to 300 MeV region leads to a modest reduction in the NN cross-section, especially for $^4$He, and we chose to the use the nuclear density dependent correction developed by Tripathi et al. [22].

For the $^4$He form factor we considered a Gaussian model with a matter radius of 1.37 fm, and also consider more a complicated form factor [6], which reproduced the minimum in the charge form factor near q$^2$ = 10 fm$^{-2}$, however a very small difference is found comparing the different form factor models and the Gaussian model is assumed for calculations below.

## 3. Results and Discussion

In **Figure 2** we show energy dependent total, elastic and absorption cross sections compared to experimental data [25-42] in the Glauber IPM and the large mass (A) limit of this model. For calculations we use Coulomb trajectories and the medium modified NN total cross sections [22], however these are found to be small corrections. The IPM and large mass limit models are in good agreement for beam energies near 1,000 MeV/u or above, however the Glauber IPM model offers improved agreement at the lower energies. For the absorption cross section comparisons, we show the data from Webber et al. [1,36] separately since here values are based on mass changing (ΔA) cross sections, which is distinct from the other measurement approaches. These data [36] are somewhat lower than the other experiments, while the models favor the higher values. The predictions of the total elastic cross sections are lower than the measurements near 1000 MeV/u, which is possibly due to the need to include correlation effects in the scattering model [19,43,44], however agree well with the measurements at higher and lower energies.

For calculations of the final state interaction (FSI), the energy deposited distribution varies with the impact parameter as shown in Figure 3. The probability to exceed 8 MeV in the energy deposition decreases from 0.37 at 1 fm to 0.26 at fm for the A=3 nuclei. To simplify the calculation, we use the values at the peak in the abrasion cross section, which occurs at impact parameter, b~2.5 fm at 100 MeV/u and b~2.0 fm at 10,000 MeV/u. **Figure 3** show how the distribution of energy deposited varies with the impact parameter. For the dominant impact parameter region of the knockout (abrasion) cross section, the final state interaction leads to a reduction in the cross section of 0.7. This value is higher than the 0.5 reduction used in the abrasion model applied to heavier nuclei [13], which should be expected because of the smaller pre-fragment nuclei considered.

**Figure 4** compares the one nucleon abrasion cross section using NN parameters averaged over neutrons and protons in interactions with the proton target. The Glauber IPM and large A limit are in good agreement, however larger differences are found for proton and neutron abrasion as discussed next. For considering $^3$He and $^3$H production cross sections we note that under the assumption that $\sigma_{nn} \sim \sigma_{pp}$, the main differences in the cross sections would be Coulomb differences in final state interactions, possible differences in the n and p internal wave functions, and the difference in $\sigma_{np}$ and $\sigma_{pp}$ in the participant and spectator factor in equation (14). Here for $^3$He production the np interaction occurs in the participant factor, and 2pp and 1np interactions occur in the spectator factor. For $^3$H production, the pp interaction occurs in the participant factor and 2np and 1pp interactions in the spectator factor. Results are shown in **Figure 5** using the FSI reduction of 0.7 applied to both $^3$He and $^3$H production cross sections. The comparisons of model to experiment show fairly good agreement, and are of similar agreement as those found comparing the different measurements reported. The predicted difference between $^3$He and $^3$H cross sections below about 200 MeV/u is large, however there is an absence of measurements to further understand this difference. One area of improvement to be considered is the role of cluster abrasion of $^3$He or $^3$H where they act as participating rather than spectators, including quantum interference effects between nucleon and cluster knockout that were predicted in our prior work [6].

We have considered the large mass limit of the Glauber model in the IPM, however it should be noted that application of the Eikonal approximation to the nucleus-nucleus multiple scattering model [45,12] leads to the identical functional form without any considerations of large A values. The first-order optical model in the Eikonal approximation for absorption and abrasion cross section relies solely on the imaginary component of the optical potential. The product forms of the cross section in the Glauber IPM model includes some contributions from both real and imaginary terms or a dependence on the $\alpha_{NN}$ values, which likely is the origin of the differences found in calculations.

In future work we will consider the deuteron production cross sections in $^4$He fragmentation, where the FSI contribution described above for $^3$He and $^3$H pre-fragments will contribute along with the abrasion of 2 nucleons, and the cluster abrasion of a deuteron should also be the focus of model building. To extend the model to lower energies (<100 MeV) additional considerations of nuclear medium effects, and classical trajectories can be considered [46]. The results here suggest $^4$He fragmentation in heavy ion collisions will likely be accurate in the abrasion model, and will be considered in future work. In addition, models of cluster abrasion will be considered in a related approach.

**Figure 1**. Right panels are results for the model fits to the total neutron-proton and proton-cross sections versus laboratory energy with experimental data [21]. The left panels show fits for neutron-proton and proton-proton scattering of the ratio of the real to imaginary parts of the NN amplitude with experimental data [22] and model estimates from Grein [22].

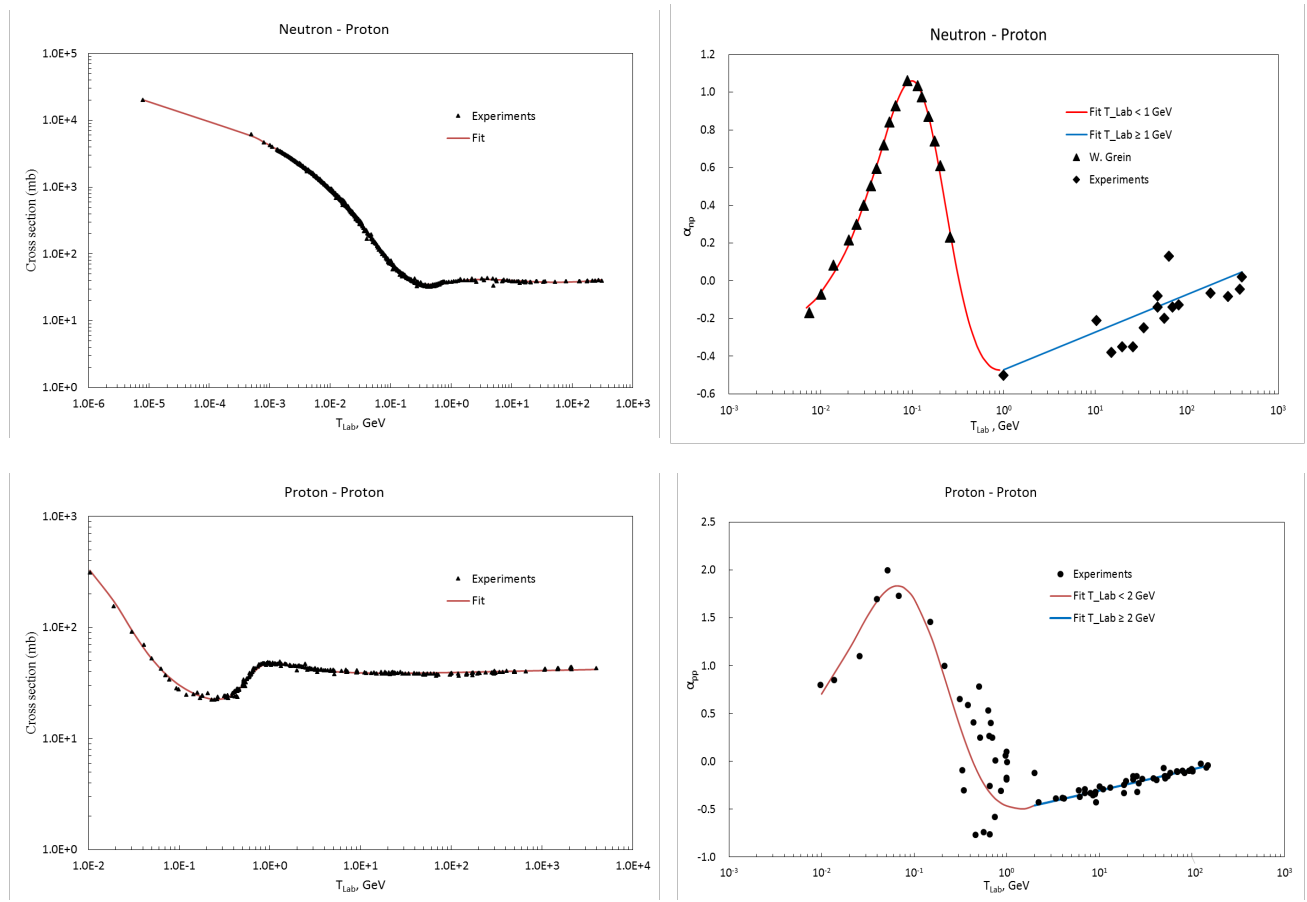

**Figure 2**. Predictions of the Glauber IPM model and large mass A limit to experimental data [25-42] for energy dependent total, absorption and elastic cross sections.

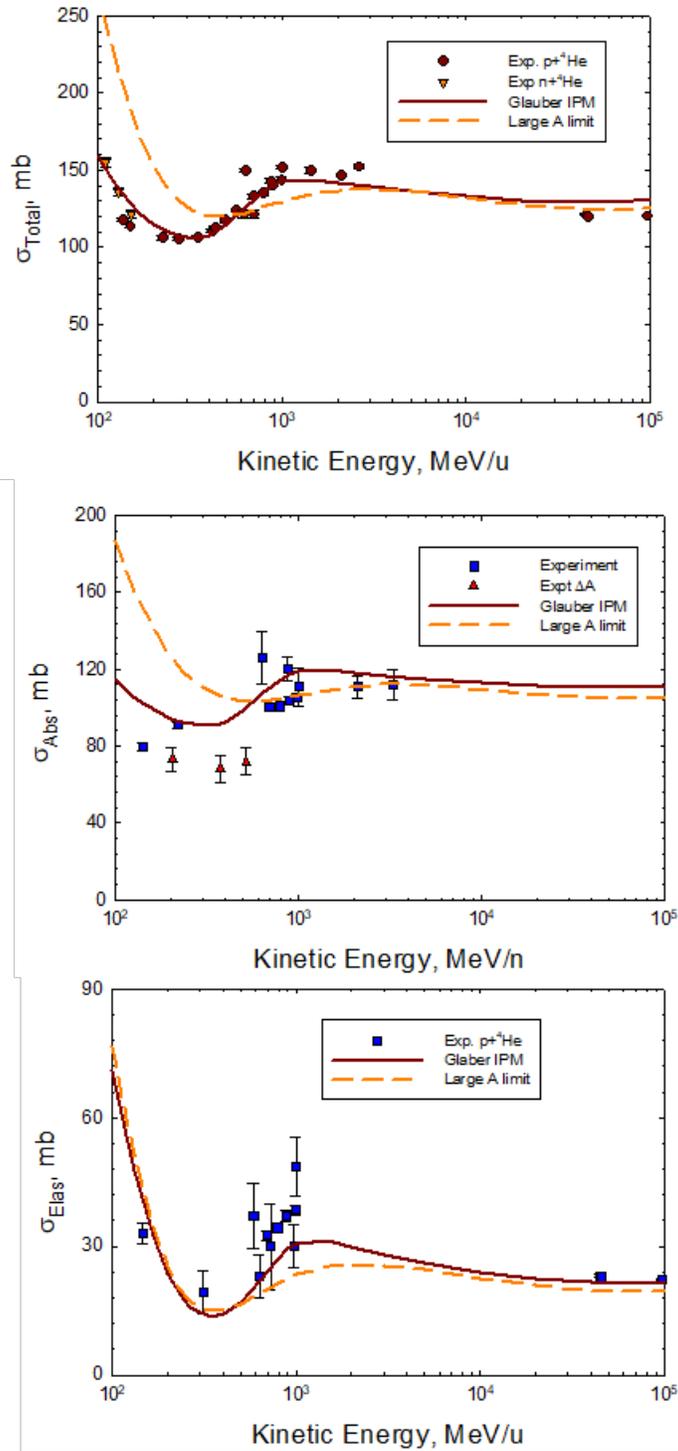

**Figure 3**. Model results for the FSI energy deposited in nucleon interactions with A=3 pre-fragments for several impact parameter values. The fraction of energy below a threshold for nucleon emission from a pre-fragment of 8 MeV for b=1, 2, 3, and 4 fm is found as 0.66, 0.7, 0.72, and 0.73, respectively.

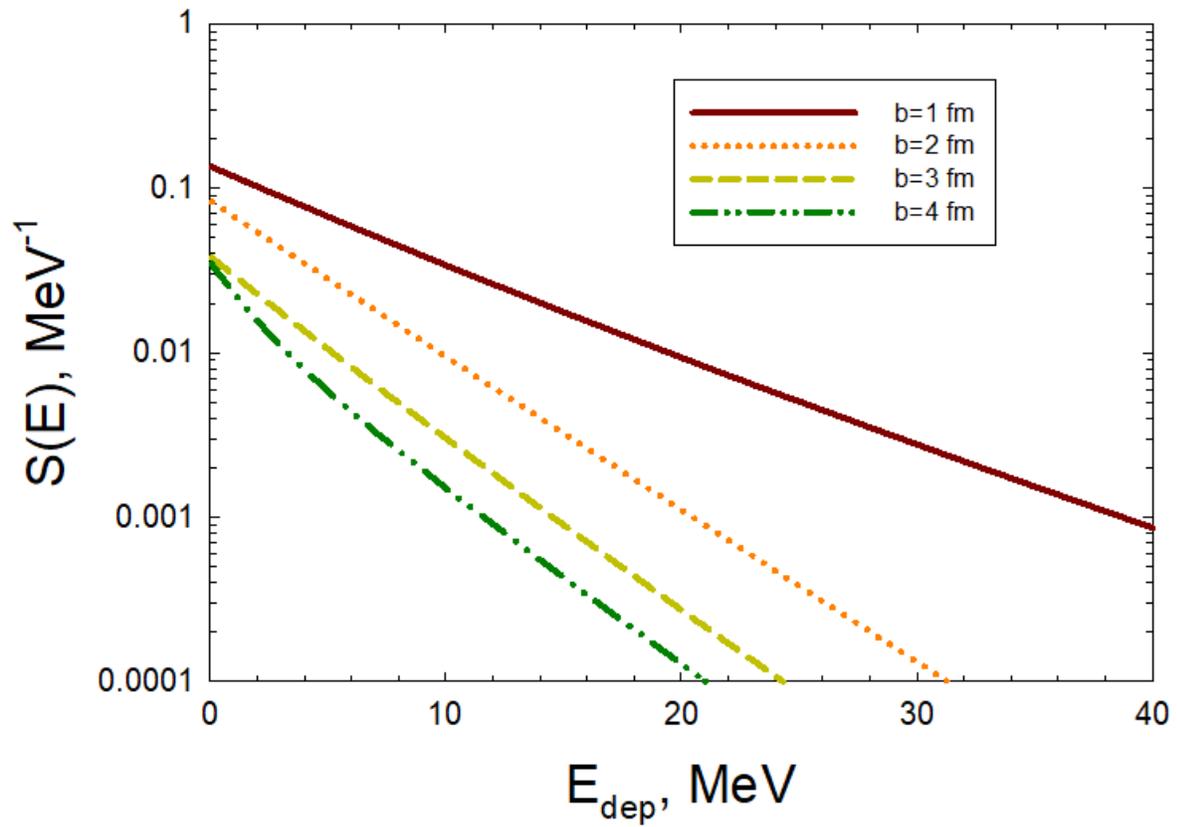

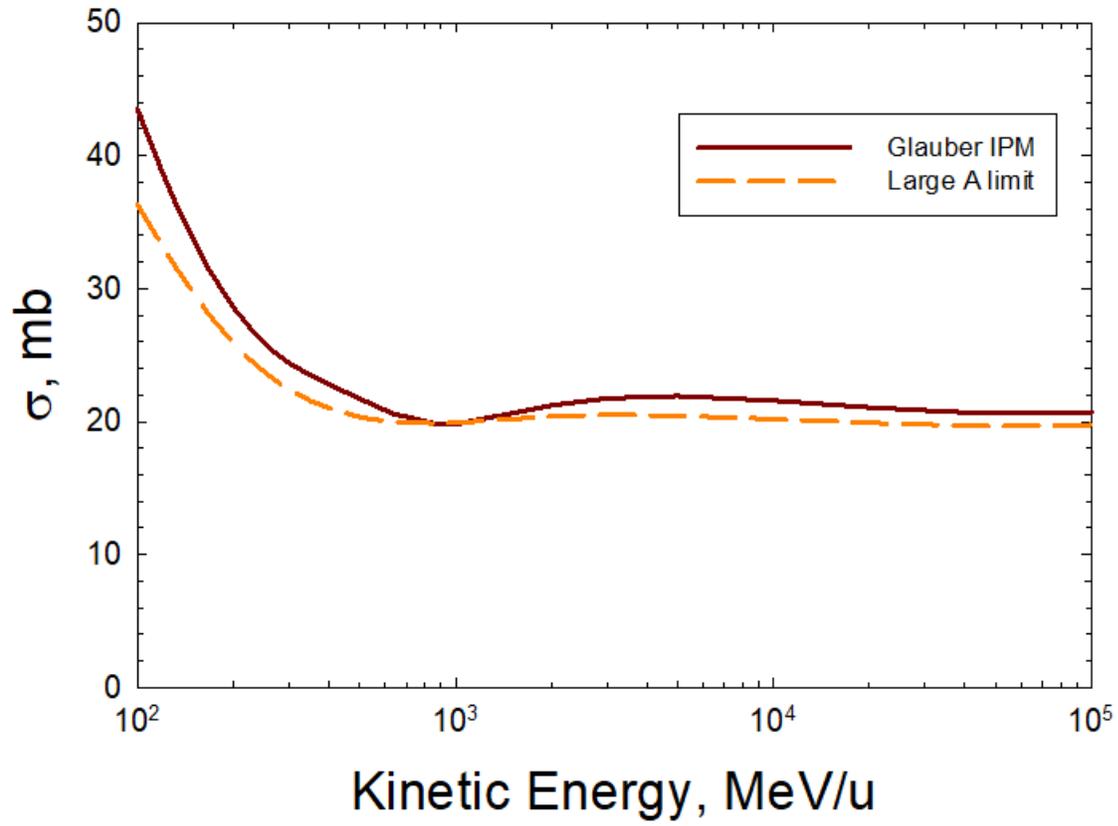

**Figure 4**. Model predictions for the energy dependence of the A=3 production cross section (without FSI).

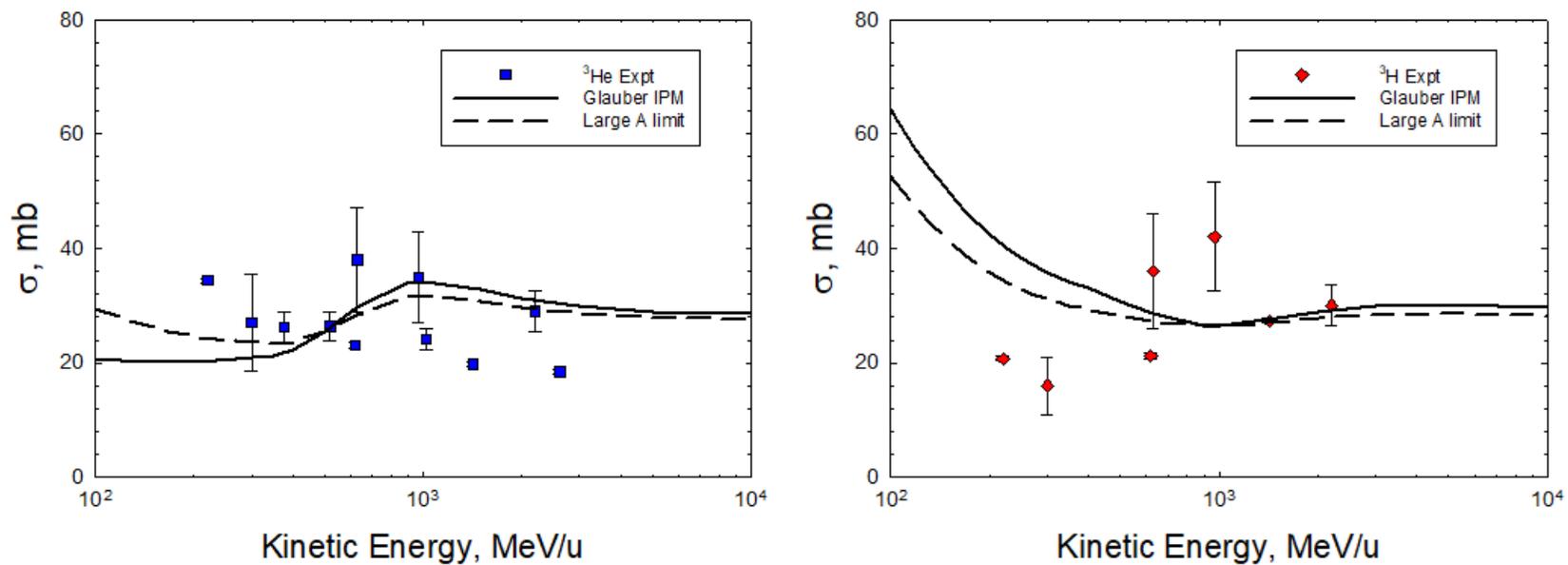

**Figure 5**. Comparison of model predictions for $^3$He (left panel) and $^3$H (right panel) in 4He-proton interactions compared to experimental data [25, 35-42].